\documentstyle[12pt]{article}

\catcode`\@=11

\@addtoreset{equation}{section}

\def\@normalsize{\@setsize\normalsize{15pt}\xiipt\@xiipt
\abovedisplayskip 14pt plus3pt minus3pt%
\belowdisplayskip \abovedisplayskip
\abovedisplayshortskip  \z@ plus3pt%
\belowdisplayshortskip  7pt plus3.5pt minus0pt}

\def\small{\@setsize\small{13.6pt}\xipt\@xipt
\abovedisplayskip 13pt plus3pt minus3pt%
\belowdisplayskip \abovedisplayskip
\abovedisplayshortskip  \z@ plus3pt%
\belowdisplayshortskip  7pt plus3.5pt minus0pt
\def\@listi{\parsep 4.5pt plus 2pt minus 1pt
            \itemsep \parsep
            \topsep 9pt plus 3pt minus 3pt}}

\def\underline#1{\relax\ifmmode\@@underline#1\else
        $\@@underline{\hbox{#1}}$\relax\fi}
\@twosidetrue

\begin{document}

\begin{flushright}
OU-HET 184\\December 1993\end{flushright}
\vspace{0.5in}

\begin{center}\Large{\bf Wheeler-DeWitt Equation \\in Two-dimensional
Quantum Gravity with $c=1$ Conformal Matter Field}\\
\vspace{3cm}
\normalsize\ Atushi Ishikawa\footnote[1]{
        e-mail: ishikawa@oskth.kek.jp}

\vspace{0.5in}
        Department of Physics \\
        Osaka University, Toyonaka, Osaka 560, Japan\\

\vspace{1.5cm}
\end{center}
\vspace{0.3in}
\baselineskip 17pt
\begin{abstract}
In the context of two-dimensional quantum cosmology,
we consider the path-integral of a string on annulus
which contains the Liouville field and
conformal matter fields.
We show that, in the transition amplitude of the string universe,
the non-zero modes of the fields are all cancelled out
only when we take the $c=1$ conformal matter field
and impose the Neumann boundary condition on the system.
The transition amplitude obtained
obeys the minisuperspace Wheeler-DeWitt equation.
In our treatment, the modular parameter on annulus
plays the role of time variable to integrate out.
\end{abstract}
\newpage

\baselineskip 20pt

\section{Introduction}

It is well-known that string theory can be viewed as two-dimensional
gravity.
Here we want to consider it as a toy model for four-dimensional
quantum cosmology.

In four dimensions, a quantum state of universe is described
by a wave function of universe on superspace (see for example
Ref.\cite{Hall}).
Superspace has a finite number of coordinates
at every point on the three-hypersurface and is infinite-dimensional.
In order to discuss quantum mechanical properties of universe, many
people want
to solve zero-energy Schr\"{o}dinger equations that are decomposed
into
the Wheeler-DeWitt equation and the momentum constraints
by the (3+1) decomposition consisting of the lapse and shift
functions ($N$, $N^i$)
and the three-metric on a hypersurface ($h_{ij}$).
It is very hard, however, to solve such infinite-dimensional
differential equations
without any approximation apart from many difficult conceptual
problems.
In order to make this problem tractable, we often reduce the
degrees of freedom of the superspace to a finite number by assuming a
symmetry.
This reduced finite dimensional superspace is called minisuperspace.
In the minisuperspace, the transition amplitude of universe
defined by the path-integral can be easily shown to obey the
Wheeler-DeWitt equation
in the $\dot{N}=0$ gauge \cite{Hall2}.

In contrast to the above four-dimensional case,
the Einstein action becomes a topological number in two dimensions.
In order to obtain a gravitational theory, we have to treat an
anomaly of
the path-integral measure exactly (see for example Ref.\cite{DHok}).
It has been proposed that the minisuperspace represents
the superspace exactly in two dimensions \cite{Moor}, \cite{Seib}.
The reason may be that an argument of a wave function of
one-dimensional universe
may be the length of the universe itself.
This conjecture is partially supported by the calculations of the
$c=0$ matrix model
\cite{Moor}.

In this paper, in the context of quantum cosmology,
we want to investigate this proposal
in the framework of the continuum Liouville theory.
In $\S$2,
in order to investigate the lapse, shift and the Liouville field
in two-dimensional gravity generally,
we start from the Polyakov action and briefly review the known
formulation
to rewrite it by using the (1+1) decomposition \cite{Teit}.
Some of this discussion in what follows is developed in
Ref.\cite{Ishikawa}.
We show that the path-integral measure $[{\cal D} g_{a b}]$ can be
decomposed into
$[{\cal D} N][{\cal D} M][{\cal D} \phi]$ by using the lapse and
shift functions
($N$, $M$) and the Liouville field ($\phi$).
There is an undesirable term consisting only of
$N$ and $M$ in the Liouville action however.

Next, in $\S$3, we take a conformal gauge in order to eliminate the
problematic term
above.
We restrict ourselves to the case of annulus topology
and calculate the transition amplitude of the one-dimensional loop
universe.
We show that
the non-zero modes of the fields are all cancelled out in the
transition amplitude
only when we take the $c=1$ conformal matter field
and impose the Neumann boundary condition on the system.
The reason is as follows.
When the cosmological constant is ignored, the Liouville field
acts as an extra conformal matter field.
Therefore a string on the worldsheet cannot vibrate
in the two-dimensional target space.
The rational for ignoring the cosmological constant
will be discussed later in this section.
This cancellation of the non-zero modes is similar to the case of
torus topology
which was investigated by Bershadsky and Klebanov \cite{Bers}.
Our guiding principle is that ghost fields should not appear on the
boundary
in the context of quantum cosmology, and this requirement is
satisfied
only by the Neumann boundary condition.
In open string theory, the Neumann boundary condition is taken,
because the end points of a string are free.
In our case, the Neumann boundary condition means that we must sum
over all
allowed values of the non-zero modes in
the initial and final states.
As a result of these settings, we obtain the transition amplitude
constructed
only by the zero modes.
It can be said that the Wheeler-DeWitt equation, that this transition
amplitude obeys,
exists only on the minisuperspace, and therefore the minisuperspace
represents
the superspace exactly.
Our result crucially depends on the fact that we consider the case
with the $c=1$ conformal matter field.


\section{The Lapse, Shift and the Liouville Field in Two-dimensional
Gravity}

In this section, we will briefly review the formulation of
two-dimensional gravity
(see for example Ref.\cite{DHok}).
Starting from the Polyakov action,
we will rewrite it by the (1+1) decomposition.
We will obtain the transition amplitude of the sting universe,
which will be used in $\S$3.

We write the coordinates on the worldsheet as
$
\xi^a=(\xi^0, \xi^1).
$
In this coordinate, the transition amplitude is
represented as
\begin{eqnarray}
Z[X^{\mu}_F, \phi_F; X^{\mu}_I, \phi_I]
      = \int \frac{[{\cal D} g_{a b}] [{\cal D} X^{\mu}]}{vol({\rm
Gauge})}
             \exp \left\{ -\frac{1}{2} \int d^2 \xi \sqrt{g} g^{a b}
			            \partial_a X^{\mu} \partial_b
X_{\mu}   \right\},
\label{ta}
\end{eqnarray}
where the metric on the worldsheet is given by
\begin{eqnarray}
ds^2 = g_{a b}(\xi) d\xi^a d\xi^b = e^{\phi(\xi)}  \hat{g}_{a b}(\xi)
d\xi^a d\xi^b.
\end{eqnarray}
Here we take a conformal time on the worldsheet, and
can take out the dynamical variable as an overall conformal mode.
We also parametrize the fiducial metric $\hat{g}_{a b}(\xi)$ by using
the lapse and shift functions
$N(\xi)$, $M(\xi)$, following the ADM decomposition in two dimensions
\cite{Teit};
\begin{eqnarray}
\hat{g}_{a b}(\xi)
  =\left( \begin{array}{cc}
     N(\xi)^{-2} + M(\xi)^2 & M(\xi) \\
     M(\xi)                 & 1
     \end{array} \right) .
\label{1+1}
\end{eqnarray}
The reason why we take an inverse of the lapse function will be
discussed later.
The most general local metric on deformations $\delta g_{a b}$ of the
metric
is given by
$
||\delta g||^2 = \int d^2 \xi \sqrt{g} \Big\{G^{a b c d} + u g^{a b}
g^{c d}
                                       \Big\} \delta g_{a b} \delta
g_{c d},
$
where $u$ is an arbitrary positive real number and $G^{a b c d}$ is
the identity
operator in the space of symmetric traceless tensors.
The decomposition of the measure $[{\cal D} g_{a b}]$ in the
transition amplitude
(\ref{ta}) is given by the orthogonal decomposition;
$
\delta g_{a b} = \delta h_{a b} + (\delta \rho) g_{a b},
$
where
\begin{eqnarray}
\delta \rho = -\frac{\delta N}{N} + \delta \phi
\end{eqnarray}
and
\begin{eqnarray}
\delta h_{a b}
  = e^{\phi} \left( \begin{array}{cc}
     \left(\frac{M^2}{N} - \frac{1}{N^3} \right) \delta N + 2M \delta
M
	 &  \frac{M}{N} \delta N + \delta M \\
     \frac{M}{N} \delta N + \delta M
	 & \frac{\delta N}{N}
     \end{array} \right).
\end{eqnarray}
Here $\delta \rho$ is the trace part of the metric deformations
$\delta g_{a b}$,
and $\delta h_{a b}$ is the symmetric traceless part.
Then the metric on deformations of $\delta g_{a b}$ is decomposed as
\begin{eqnarray}
||\delta g||^2 = \int d^2 \xi \sqrt{g} G^{a b c d} \delta h_{a b}
\delta h_{c d}
                   + 4u \int d^2 \xi \sqrt{g} (\delta \rho)^2.
\end{eqnarray}
{}From this decomposition, we can separate the measure $[{\cal D} g_{a
b}]$
in the form of
$
[{\cal D} \rho] [{\cal D} h_{a b}]
$.
Next we change variables from $\rho$, $h_{a b}$ to $\phi$, $v_a$;
$
\delta \rho = \delta \phi + g^{a b} \bigtriangledown_a (\delta v_b)
$,
$\delta h_{a b} = 2 G^{c d}_{a b} \bigtriangledown_c (\delta v_d)
                = \left( P_1 \delta v \right)_{a b},
$
where $\delta v_a$ are infinitesimal generators with respect to a
two-dimensional
diffeomorphism.
The operator $P_1$
maps vectors into symmetric traceless tensors.
We obtain another decomposition with the Jacobian
$\left\{det^{\prime}(P_1^{\dagger} P_1) \right\}^{1/2}_{g} $;
\begin{eqnarray}
[{\cal D} g_{a b}]
           = \prod_{i} d \tau_i \frac{1}{vol({\rm CK})}
		         \frac{<\psi^{(j)}|\frac{\partial g}{\partial
\tau_k}>_g}
			       {<\psi^{(j)}|\psi^{(k)}>^{1/2}_g}
		         [{\cal D} \phi] [{\cal D} v_{a}]
				    \left\{det^{\prime}(P_1^{\dagger}
P_1) \right\}^{1/2}_{g},
\label{ippan}
\end{eqnarray}
where $\tau_i$ are  moduli parameters and the factor
$\frac{<\psi^{(j)}|\frac{\partial g}{\partial \tau_k}>_g}
{<\psi^{(j)}|\psi^{(k)}>^{1/2}_g}$
is the Weil-Petersson measure which represents the angle between the
moduli space
and the gauge orbit of a diffeomorphism.
The $prime$ in equation (\ref{ippan}) denotes the omission of
zero-mode with respect to conformal Killing vectors $dV$;
$P_1(dV) = 0$.
We must divide by the volume of conformal Killing vectors $vol({\rm
CK})$.
The Liouville action $S_{\phi}$ arises in the formula;
\begin{eqnarray}
  		         \frac{<\psi^{(j)}|\frac{\partial g}{\partial
\tau_k}>_g}
			       {<\psi^{(j)}|\psi^{(k)}>^{1/2}_g}
    \left\{det^{\prime}(P_1^{\dagger} P_1) \right\}^{1/2}_{g}
   =
		   	    \frac{<\psi^{(j)}|\frac{\partial
\hat{g}}{\partial \tau_k}>_{\hat{g}}}
		        {<\psi^{(j)}|\psi^{(k)}>^{1/2}_{\hat{g}}}
	\left\{det^{\prime}(P_1^{\dagger} P_1)
\right\}^{1/2}_{\hat{g}}
	        e^{-\frac{26}{48\pi} S_{\phi}[\phi, \hat{g}]}.
\end{eqnarray}
Accordingly the measure is decomposed as
\begin{eqnarray}
[{\cal D} g_{a b}]
           &=& \prod_i d \tau_i
		   	    \frac{<\psi^{(j)}|\frac{\partial
\hat{g}}{\partial \tau_k}>_{\hat{g}}}
		        {<\psi^{(j)}|\psi^{(k)}>^{1/2}_{\hat{g}}}
                [{\cal D} \phi] [{\cal D} v_{a}]
				      \left\{det^{\prime} (P_1^{\dagger} P_1)
\right\}^{1/2}_{\hat{g}}
                      e^{-\frac{26}{48\pi} S_{\phi}[\phi, \hat{g}]}
\nonumber \\*
           &\equiv& [{\cal D} \phi] [{\cal D} \hat{h}_{a b}]
		              e^{-\frac{26}{48\pi} S_{\phi}[\phi,
\hat{g}]},
\label{decomposition}
\end{eqnarray}
where $\hat{h}_{a b}$ is defined by above equation.
The most general local metric on deformations  $\delta \hat{h}_{a b}$
is given by
\begin{eqnarray}
||\delta \hat{h}||^2 &=& \int d^2 \xi
           \sqrt{\hat{g}} \left\{\hat{G}^{a b c d}
		           + u \hat{g}^{a b} \hat{g}^{c d}
                        \right\} \delta \hat{h}_{a b} \delta
\hat{h}_{c d}
\nonumber \\*
          &=& \int d^2 \xi \left\{ 2 \frac{(\delta N)^2}{N^2}
		                             + 2 N^2 (\delta M)^2
\right\}.
\end{eqnarray}
{}From this decomposition, we obtain another separation of the measure
$[{\cal D} g_{a b}]$;
\begin{eqnarray}
[{\cal D} g_{a b}]
           =  [{\cal D} \phi] [{\cal D} N] [{\cal D} M]
		              e^{-\frac{26}{48\pi} S_{\phi}[\phi,
\hat{g}]}.
\label{de}
\end{eqnarray}
In obtaining this form, we have taken the parametrization of the
fiducial metric
given by equation (\ref{1+1}).
Here we find the same separation form of the measure that we take in
four dimensions.
This discussion is developed in Ref.\cite{Ishikawa}.
We should be careful about treating the measure of the Liouville
field $[{\cal D} \phi]$,
because it is not translationally invariant. We will discuss it in
$\S$3.

The difference of the path-integral for conformal matter fields
(central charge $c$) evaluated on $g_{a b}$ and that
on $\hat{g}_{a b}$ is also represented
in terms of the Liouville action.
Consequently the transition amplitude (\ref{ta}) can be expressed as
\begin{eqnarray}
Z[X^{\mu}_F, \phi_F; X^{\mu}_I, \phi_I]
       &=& \int \frac{[{\cal D} N] [{\cal D} M]}{vol({\rm Gauge})}
	     \int [{\cal D} X^{\mu}]
             \exp \left\{ -\frac{1}{2} \int d^2 \xi \sqrt{\hat{g}}
\hat{g}^{a b}
			            \partial_a X^{\mu} \partial_b
X_{\mu}   \right\}
\nonumber \\*
&&\times \int [{\cal D} \phi]
						e^{-\frac{26-c}{48\pi}
S_{\phi}[\phi, \hat{g}]}.
\label{ta2}
\end{eqnarray}
By using the parametrization of the fiducial metric (\ref{1+1}),
we can write the action of conformal matter fields as
\begin{eqnarray}
S_{X^{\mu}}       &=& \frac{1}{2} \int d^2 \xi \sqrt{\hat{g}}
\hat{g}^{a b}
            \partial_a X^{\mu} \partial_b X_{\mu}
\nonumber \\*
		  &=& \int d^2 \xi \left\{
		     {P_X}_{\mu} {\dot{X}}^{\mu} - N^{-1}
H^{X^{\mu}}_0 - M H^{X^{\mu}}_1
				             \right\},
\end{eqnarray}
where
\begin{eqnarray}
{P_X}^{\mu} &=& N \left( {\dot{X}^{\mu}} - M {X^{\mu}}^{\prime}
\right),
\nonumber \\*
H^{X^{\mu}}_0 &=& \frac{1}{2} \left( {P_X}^{\mu} {P_X}_{\mu}
                           - {X^{\mu}}^{\prime} {X_{\mu}}^{\prime}
\right),
{}~~~H^{X^{\mu}}_1 = {P_X}_{\mu} {X^{\mu}}^{\prime}.
\end{eqnarray}
Likewise,
\begin{eqnarray}
S_{\phi}          &=& \kappa \int d^2 \xi \sqrt{\hat{g}}
                       \left\{
					\frac{1}{2} \hat{g}^{a b}
\partial_a \phi \partial_b \phi
                   + \hat{R} \phi
				   + \mu e^{\phi}
					   \right\}
\nonumber \\*
				  &=& \int d^2 \xi  \left\{
				                 P_{\phi} \dot{\phi}
- N^{-1} H^{\phi}_0 - M H^{\phi}_1
	+ 2\kappa N {M^{\prime}}^2
	  \right\},
\label{la}
\end{eqnarray}
where
\begin{eqnarray}
P_{\phi} &=& \kappa N \left( \dot{\phi}
                   - M \phi^{\prime} - 2M^{\prime} \right),
\nonumber \\*
H^{\phi}_0 &=& \frac{1}{2 \kappa} P_{\phi}^2 -
\frac{\kappa}{2}{\phi^{\prime}}^2
                   + 2 \kappa \phi^{\prime \prime} + \mu e^{\phi},
{}~~~H^{\phi}_1 = P_{\phi} \phi^{\prime} - 2\phi^{\prime}.
\end{eqnarray}
The $dot$ and the $prime$ represent the derivative with respect to
$\xi^0$
and $\xi^1$ respectively, and we take $\kappa=\frac{26-c}{48 \pi}$.

The extra term $2\kappa N {M^{\prime}}^2$ in the Liouville action
(\ref{la})
prevents us from interpreting the lapse and shift functions as
the Lagrange multipliers \cite{Teit}, \cite{Ishikawa}, \cite{Yone}.
This is related to the fact that there is
no invariance under the diffeomorphisms of the fiducial metric
$\hat{g}_{a b}$
where the Liouville mode $\phi$ is taken out.
Therefore we cannot construct the canonical formulation of this
system;
the degrees of freedom are insufficient and there is no canonical
structure (classically).
In order to avoid this problem,
Teitelboim added an extra field to the system in order to recover the
canonical
structure \cite{Teit}, \cite{Yone}, \cite{Mart}.
The meaning of this extra field is unclear however.
In the next section, we will bypass this problem for the case of
annulus topology.


\section{The Transition Amplitude for the case of annulus}

In this section, we concentrate on the case of annulus topology
and fix the diffeomorphisms completely by taking
the easiest gauge, namely the conformal gauge,
in order to eliminate the problematic term mentioned in the previous
section.
This gauge fixing is quite different from the $\dot{N} =0$ gauge used
in
the minisuperspace treatment of a four-dimensional gauge fixing
(see for example Ref.\cite{Hall}\cite{Hall2}).
If we fix the diffeomorphisms completely, this problematic term
disappears
and the degrees of freedom in the fiducial metric $\hat{g}_{a b}$
become finite, and is described by a modular parameter $t$
in the annulus case.
In other words, we can take $N^{-1}=t$ and $M=0$ in (2.3).
In this gauge, from the decomposition of the measure $[{\cal D} g_{a
b}]$
given by the first line of equation (\ref{decomposition}),
we obtain the transition amplitude of the one-dimensional loop
universe;
\begin{eqnarray}
Z[X^{\mu}_F, \phi_F; X^{\mu}_I, \phi_I] &=&
      \int_0^{\infty} dt \frac{1}{\Omega(CK)}
              \frac{<\psi|\frac{\partial \hat{g}}{\partial
t}>_{\hat{g}}}
			       {<\psi|\psi>^{1/2}_{\hat{g}}}
			  \left\{det^{\prime}(P_1^{\dagger} P_1)
\right\}^{1/2}_{\hat{g}}
\nonumber \\*
	&&\times \int [{\cal D} X^{\mu}] e^{-S_{X^{\mu}}[X^{\mu}, t]}
	         \int [{\cal D} \phi] e^{-S_{\phi} [\phi, t]}
\nonumber \\*
    &=&      \int_0^{\infty}    \frac{dt}{t}
			       \left\{det^{\prime}(P_1^{\dagger} P_1)
\right\}^{1/2}_{\hat{g}}
	       \int [{\cal D} X^{\mu}] e^{-S_{X^{\mu}}[X^{\mu}, t]}
	       \int [{\cal D} \phi] e^{-S_{\phi} [\phi, t]},
\label{ta3}
\end{eqnarray}
where
we have used the following results of the calculations
of the Weil-Petersson measure and the volume of a conformal Killing
vector;
\begin{eqnarray}
              \frac{<\psi|\frac{\partial \hat{g}}{\partial
t}>_{\hat{g}}}
			       {<\psi|\psi>^{1/2}_{\hat{g}}}
             = \left( \frac{2}{t} \right)^{1/2}
\end{eqnarray}
and
\begin{eqnarray}
\Omega(CK)  = t^{1/2}.
\end{eqnarray}
Note that the volume of $v_a$ is divided by $vol({\rm Gauge})$.
Here the action of the conformal matter fields and
that of the Liouville field can be rewritten respectively
as
\begin{eqnarray}
S_{X^{\mu}}[X^{\mu}, t] =  \frac{1}{2} \int_M d^2 \sigma
		    \left\{ \dot{X^{\mu}} \dot{X_{\mu}}
			       + {X^{\mu}}^{\prime}
{X_{\mu}}^{\prime} \right\}
\label{ca}
\end{eqnarray}
and
\begin{eqnarray}
S_{\phi}[\phi, t] = \frac{\kappa}{2} \int_M d^2 \sigma
		    \left\{ \dot{\phi}^2 + {\phi^{\prime}}^2
			   - 4 \phi^{\prime \prime} \right\}.
\label{la2}
\end{eqnarray}
Here we have simply discarded the cosmological constant
(see the discussion later for the case of $c=1$).

In order to obtain the actions (\ref{ca}) and (\ref{la2}),
we have made a coordinate transformation from
$\xi^a$ to $\sigma^a$ $(\sigma^0 = t \xi^0,~ \sigma^1 = \xi^1 )$
and the region $M$ of the new coordinates $\sigma^a$ is given by
\begin{eqnarray}
M:~~~0 \le \sigma^0 \le t,~~~0 \le \sigma^1 \le 1.
\end{eqnarray}
Here the space coordinate $\sigma^1$ is periodic
and the boundaries exist at $\sigma^0 = 0, t$.
{}From this transformation,
we can interpret the modular parameter $t$ as a time variable
of the system consisting of
conformal matter fields and the Liouville field.

For the computation of the transition amplitude (\ref{ta3}),
we will make the mode expansion.
For conformal matter fields, we expand them as follows
\begin{eqnarray}
X^{\mu}(\sigma^0, \sigma^1)
    = X^{\mu}_0(\sigma^0) + \sum_{n \not= 0} a^{\mu}_n(\sigma^0)
e^{-2\pi i \sigma^1}.
\end{eqnarray}
The first term is the zero mode and the second term represents the
vibrations.
We can separate the partition function of matter fields
into the zero mode part and the non-zero modes parts;
\begin{eqnarray}
\int [{\cal D} X^{\mu}] e^{-S_{X^{\mu}} [X^{\mu}, t]}
  &=& \int [{\cal D} X^{\mu}_0] e^{- S_{X^{\mu}_0}[X^{\mu}_0, t]}
        \prod_{n \not= 0} \int  [{\cal D} a^{\mu}_n]
	         e^{- S_{a^{\mu}_n}[a^{\mu}_n, t]}
 \nonumber \\*
 &=& K(X^{\mu}_{0 F}, t; X^{\mu}_{0 I}, 0)
        \prod_{n \not= 0} \left\{ \frac{1}{n {\rm sinh}(2\pi n t)}
\right\}^{c/2},
\label{cp}
\end{eqnarray}
where
\begin{eqnarray}
S_{X^{\mu}}[X^{\mu}, t] &=& S_{X^{\mu}_0}[X^{\mu}_0, t]
                             + \sum_{n \not= 0}
S_{a^{\mu}_n}[a^{\mu}_n, t],
\nonumber \\*
S_{X^{\mu}_0}[X^{\mu}_0, t]
                &=& \frac{1}{2} \int^t_0 d \sigma^0 \dot{X}^{\mu}_0
\dot{X}_{0 \mu},
\nonumber \\*
S_{a^{\mu}_n}[a^{\mu}_n, t] &=& \frac{1}{2} \int^t_0 d \sigma^0
                      \left\{ {\dot{a}}^{\mu}_n {\dot{a}}_{n \mu}
					       + (2\pi n)^2 a^{\mu}_n
a_{n \mu} \right\},
\label{cmea}
\end{eqnarray}
and
\begin{eqnarray}
K(X^{\mu}_{0 F}, t; X^{\mu}_{0 I}, 0) &\equiv&
\int [{\cal D} X^{\mu}_0] e^{- S_{X^{\mu}_0}[X^{\mu}_0, t]}
\nonumber \\*
&=&  \left( \frac{1}{2\pi t} \right)^{c/2}
     \exp \left\{ - \frac{ (X^{\mu}_{0 F} - X^{\mu}_{0 I})
	                       (X_{0 F \mu} - X_{0 I \mu})   }{2t}
\right\},
\nonumber \\*
        \int  [{\cal D} a^{\mu}_n] e^{- S_{a^{\mu}_n}[a^{\mu}_n, t]}
                &=& \left\{ \frac{1}{n {\rm sinh}(2\pi n t)}
\right\}^{c/2}.
\end{eqnarray}
We use the knowledge of the path-integral in quantum mechanics
for a free particle and a harmonic oscillator.
Here we take the Neumann boundary condition as stated in the
introduction.
Note that the Neumann boundary condition puts no restriction on the
zero mode.
For the Liouville field, we also make the same mode expansion;
\begin{eqnarray}
\phi(\sigma^0, \sigma^1)
    = \phi_0(\sigma^0) + \sum_{n \not= 0} b_n(\sigma^0) e^{-2\pi i
\sigma^1}.
\end{eqnarray}
We also separate the partition function
of the Liouville field into two parts;
\begin{eqnarray}
\int [{\cal D} \phi] e^{-S_{\phi} [\phi, t]}
 =K(\phi_{0 F}, t; \phi_{0 I}, 0)
        \prod_{n \not= 0} \left\{ \frac{1}{n {\rm sinh}(2\pi n t)}
\right\}^{1/2},
\label{lp}
\end{eqnarray}
where
\begin{eqnarray}
S_{\phi}[\phi, t] &=& S_{\phi_0}[\phi_0, t] + \sum_{n \not= 0}
S_{b_n}[b_n, t],
\nonumber \\*
S_{\phi_0}[\phi_0, t] &=& \frac{\kappa}{2} \int^t_0 d \sigma^0
\dot{\phi}^2_0,
\nonumber \\*
S_{b_n}[b_n, t] &=& \frac{\kappa}{2} \int^t_0 d \sigma^0
                      \left\{ \dot{b_n}^2 + (2\pi n)^2 b_n^2
\right\}.
\label{lmea}
\end{eqnarray}
The second derivative term in the Liouville action (\ref{la2})
disappears in the mode expansion.
For the ghost fields, we also take the Neumann boundary condition.
We obtain the Faddeev-Popov determinant
\begin{eqnarray}
\left\{det^{\prime}(P_1^{\dagger} P_1) \right\}^{1/2}_{\hat{g}}
=2\pi t \prod_{n \not= 0} n {\rm {\rm sinh}}(2\pi n t).
\label{fpd}
\end{eqnarray}
There is no zero mode, because it is absent from beginning.

Inserting the partition functions of conformal matter fields and the
Liouville field
(\ref{cp}), (\ref{lp}) and the Faddeev-Popov determinant (\ref{fpd})
into equation (\ref{ta3}),
we obtain finally the transition amplitude for the case of annulus
topology explicitly;
\begin{eqnarray}
Z[X^{\mu}_F, \phi_F; X^{\mu}_I, \phi_I]
   &=& 2\pi \int_0^{\infty} dt
                 K(X^{\mu}_{0 F}, t; X^{\mu}_{0 I}, 0)
                 K(\phi_{0 F}, t; \phi_{0 I}, 0)
\nonumber \\*
&&\times	     \left\{ \prod_{n \not= 0} n {\rm {\rm
sinh}}(2\pi n t)
				 \right\}^{(1-c)/2}.
\label{ta4}
\end{eqnarray}
We can easily show that the partition function for the zero mode and
the one for the non-zero modes
satisfy the Euclidean Schr\"{o}dinger equation individually;
\begin{eqnarray}
&&\hat{H}_0^{X^{\mu}_0} K(X^{\mu}_{0 F}, t; X^{\mu}_{0 I}, 0)
               = -\frac{\partial}{\partial t} K(X^{\mu}_{0 F}, t;
X^{\mu}_{0 I}, 0),
\nonumber \\*
&&\hat{H}_0^{a^{\mu}_n} K(a^{\mu}_{n F}, t; a^{\mu}_{n I}, 0)
               = -\frac{\partial}{\partial t} K(a^{\mu}_{n F}, t;
a^{\mu}_{n I}, 0)
\label{cE}
\end{eqnarray}
and
\begin{eqnarray}
&&\hat{H}_0^{\phi_0} K(\phi_{0 F}, t; \phi_{0 I}, 0)
               = -\frac{\partial}{\partial t} K(\phi_{0 F}, t;
\phi_{0 I}, 0),
\nonumber \\*
&&\hat{H}_0^{b_n} K(b_{n F}, t; b_{n I}, 0)
               = -\frac{\partial}{\partial t} K(b_{n F}, t; b_{n I},
0).
\label{lE}
\end{eqnarray}
The Hamiltonian operators
$\hat{H}_0^{X^{\mu}_0}$, $\hat{H}_0^{a^{\mu}_n}$,
$\hat{H}_0^{\phi_0}$
and $\hat{H}_0^{b_n}$
are defined as follows. From the mode-expanded actions
(\ref{cmea}) and (\ref{lmea}), we can rewrite the actions and
Hamiltonians;
\begin{eqnarray}
S_{X^{\mu}}[X^{\mu}, t] = \int_0^t d \sigma^0 \left\{
               {P_{X_0}}_{\mu} {\dot{X}^{\mu}}_0
			   + \sum_{n \not= 0} {P_{a_n}}_{\mu}
\dot{a}^{\mu}_n - H_0^{X^{\mu}}
			   \right\}
\end{eqnarray}
and
\begin{eqnarray}
S_{\phi}[\phi, t] = \int_0^t d \sigma^0 \left\{
               P_{\phi_0} \dot{\phi}_0
			   + \sum_{n \not= 0} P_{b_n} \dot{b}_n -
H_0^{\phi}
			   \right\},
\end{eqnarray}
where
\begin{eqnarray}
H_0^{X^{\mu}} &=& \frac{1}{2} \left[ {P_{X_0}}^{\mu} {P_{X_0}}_{\mu}
+ \sum_{n \not= 0}
                            \left\{\dot{a}^{\mu}_n \dot{a}_{n \mu}
							    - (2\pi
n)^2 a^{\mu}_n a_{n \mu} \right\}
                      \right]
        \equiv H_0^{X^{\mu}_0} + \sum_{n \not= 0} H_0^{a^{\mu}_n},
\nonumber \\*
H_0^{\phi} &=& \frac{\kappa}{2} \left[ P_{\phi_0}^2 + \sum_{n \not=
0}
                                 \left\{\dot{b}_n^2 - (2\pi n)^2
b_n^2 \right\}
                      \right]
        \equiv H_0^{\phi_0} + \sum_{n \not= 0} H_0^{b_n}.
\end{eqnarray}
In order to obtain the Hamiltonian operators, we replace the momentum
by
the differential operator.

We observe that the last term of the transition amplitude (\ref{ta4})
becomes a number only when we take $c=1$.
This means that there is no vibration of the string as stated in the
introduction.
Therefore only in the case of $c=1$,
we can show that the transition amplitude of the string universe
obeys the minisuperspace
Wheeler-DeWitt equation, by using the Euclidean Schr\"{o}dinger
equations
(\ref{cE}) and (\ref{lE});
\begin{eqnarray}
\hat{H}_0 Z[X_F, \phi_F; X_I, \phi_I]
    &=& \left( \hat{H}_0^X + \hat{H}_0^{\phi} \right) Z[X_F, \phi_F;
X_I, \phi_I]
\nonumber \\*
    &=& \left( \hat{H}_0^{X_0} + \hat{H}_0^{\phi_0} \right) Z[X_F,
\phi_F; X_I, \phi_I]
\nonumber \\*
	&=& - 2\pi \int_0^{\infty} dt \frac{\partial}{\partial t}
\left\{
           K(X_{0 F}, t; X_{0 I}, 0) K(\phi_{0 F}, t; \phi_{0 I}, 0)
\right\}
\nonumber \\*
  &=& - 2\pi \left[ K(X_{0 F}, t; X_{0 I}, 0)
                 K(\phi_{0 F}, t; \phi_{0 I}, 0) \right]^{\infty}_0
\nonumber \\*
  &=& \delta (X_{0 F} - X_{0 I}) \delta (\phi_{0 F} - \phi_{0 I}).
\label{wd}
\end{eqnarray}
Here, in the case of $c=1$,
the reason why we could ignore the cosmological constant is clear
by following the discussions in Ref.\cite{Bers}\cite{Kleb};
the theory cut-off by the exponential interaction
which originates from the cosmological constant
may be identified with the free field theory with an appropriate
renormalization.
We do not have to take the modification of the Liouville field
dynamics proposed by David and by Distler and Kawai \cite{Davi},
\cite{Dist},
when we only consider the annulus topology without cosmological
constant.

As is well known, the quantity
$\prod_{n \not= 0} n {\rm {\rm sinh}}(2\pi n t)$
in the Faddeev-Popov determinant diverges. If we regularize this
quantity
by using $\zeta$ function, we obtain the well-known form
\begin{eqnarray}
\left\{det^{\prime}(P_1^{\dagger} P_1) \right\}^{1/2}_{\hat{g}}
          &=& 2\pi t \prod_{n \not= 0} n {\rm {\rm sinh}}(2\pi n t)
\nonumber \\*
          &=& 2\pi t e^{-\frac{\pi t}{3}} \prod^{\infty}_{n=1}
		      \left( 1 - e^{-4\pi n t} \right)^2.
\end{eqnarray}
This regularization is not important, because these divergent
quantities
are all cancelled out in the transition amplitude (\ref{ta4})
in the case of $c=1$.

On the other hand, if we take the Dirichlet boundary condition
for the conformal matter fields and the Liouville field,
we obtain the transition amplitude;
\begin{eqnarray}
&&Z[X^{\mu}_F, \phi_F; X^{\mu}_I, \phi_I]
\nonumber \\*
&&= 2\pi \int_0^{\infty} dt
     K(X^{\mu}_{0 F}, t; X^{\mu}_{0 I}, 0)  K(\phi_{0 F}, t; \phi_{0
I}, 0)
    \prod_{n \not= 0} n^{(3+c)/2} \left\{ \sinh(2\pi n t)
\right\}^{(1-c)/2}
\nonumber \\*
&\times& \exp \Bigg[ -\sum_{n \not= 0} \frac{\pi n}{\sinh(2\pi n)}
                 \Big\{ \cosh(2\pi n t) \left(a^{\mu}_{n F} a_{n F
\mu}
				                              +
a^{\mu}_{n I} a_{n I \mu}
				                              +
b^{\mu}_{n F} b_{n F \mu}
			  + b^{\mu}_{n I} b_{n I \mu}
				 \right)
\nonumber \\*
&&~~~~~~~~~~~~~~~~~~~~~~~~~~~~~~~~
-\left(2a^{\mu}_{n F} a_{n I \mu} + 2b^{\mu}_{n F} b_{n I \mu}
\right)
				 \Big\} \Bigg].
\label{ta5}
\end{eqnarray}
The Dirichlet boundary condition is not suited even for the $c=1$
case.
One reason is that
if we regularize the quantity $\prod_{n \not= 0} n^2$, it becomes
zero and
the transition amplitude given by equation (\ref{ta5})
becomes meaningless. The other reason is as follows.
In the Dirichlet case, we are left with
the non-zero modes of the conformal matter field and
those of the Liouville field on the boundary.
In addition, the factor $\sinh^{-1} (2\pi n)$
is cancelled out by the Faddeev-Popov determinant, and
therefore the non-zero modes on the boundary cannot satisfy the
Euclidean Schr\"{o}dinger
equations (\ref{cE}), (\ref{lE}) by using the parts which do not
cancel out.
In this case, the transition amplitude (\ref{ta5}) cannot obey
the Wheeler-DeWitt equation.
These are the reasons why we have chosen the Neumann boundary
condition.

Our result depends crucially on the fact that we have considered
the case with the $c=1$ conformal matter field.
The situation for $c \not= 1$ is beyond the scope of this paper,
because
it is impossible to cancel out the vibrations of the non-zero modes.

\section{Conclusion}

In this paper, we have pointed out that there is lack of canonical
structure in
two-dimensional gravity \cite{Ishikawa}.
In order to eliminate the problematic term,
we have taken a conformal gauge by concentrating on the case of
annulus topology.
Our main result is the transition amplitude of
the string universe (\ref{ta4})
which obeys the minisuperspace Wheeler-DeWitt equation (\ref{wd}).
We have restricted ourselves to the case of the $c=1$ conformal
matter field
and have imposed the Neumann boundary
condition.

There may be other methods to avoid this problem,
and it is worth investigating
different gauge fixing (for example Ref.\cite{Free}).

\vskip1.5cm

\noindent Acknowledgments

We would like to express our gratitude to Prof. Itoyama and Prof.
Kubota
for their helpful discussions.
Thanks are also due to Prof. Itoyama for a careful reading of the
manuscript and the other members of High
Energy Physics group in Osaka University for
their warm encouragements.

\vskip1.5cm


\end{document}